\documentclass[times,twocolumn,final]{elsarticle}

\usepackage{medima}
\usepackage{framed,multirow}

\usepackage{amssymb}
\usepackage{latexsym}

\usepackage{xcolor}
\usepackage{amsmath,graphicx}
\usepackage{gensymb}
\usepackage{enumitem}
\usepackage{url}
\usepackage{txfonts}
\usepackage{hyperref}
\usepackage{multirow}

\definecolor{newcolor}{rgb}{.8,.349,.1}

\journal{Medical Image Analysis}

\begin{document}

\verso{Pavel Sinha \textit{et~al.}}

\begin{frontmatter}

\title{CNN-based automatic segmentation of Lumen \& Media boundaries in IVUS images using closed polygonal chains}%


\author[1]{Pavel \snm{Sinha}\corref{cor1}}
\cortext[cor1]{Corresponding author: pavel.sinha@mail.mcgill.ca}
\author[1]{Ioannis \snm{Psaromiligkos}\fnref{fn1}}
\fntext[fn1]{ioannis.psaromiligkos@mcgill.ca (Ioannis Psaromiligkos)}
\author[1]{Zeljko \snm{Zilic}\fnref{fn2}}
\fntext[fn2]{zeljko.zilic@mcgill.ca (Zeljko Zilic)}

\address[1]{Department of Electrical and Computer Engineering, McGill University, Montreal, Quebec, Canada}

\received{22 September 2023}

\begin{abstract}
We propose an automatic segmentation method for lumen and media with irregular contours in IntraVascular ultra-sound (IVUS) images.
In contrast to most approaches that broadly label each pixel as either lumen, media, or background, we propose to approximate the lumen and media contours by closed polygonal chains. 
The chain vertices are placed at fixed angles obtained by dividing the entire 360\degree~angular space into equally spaced angles, and we predict their radius using an adaptive-subband-decomposition CNN.
We consider two loss functions during training.
The first is a novel loss function using the Jaccard Measure (JM) to quantify the similarities between the predicted lumen and media segments and the corresponding ground-truth image segments.
The second loss function is the traditional Mean Squared Error. 
The proposed architecture significantly reduces computational costs by replacing the popular auto-encoder structure with a simple CNN as the encoder and the decoder is reduced to simply joining the consecutive predicted points.
We evaluated our network on the publicly available IVUS-Challenge-2011 dataset using two performance metrics, namely JM and Hausdorff Distance (HD). 
The evaluation results show that our proposed network mostly outperforms the state-of-the-art lumen and media segmentation methods.
\end{abstract}


\end{frontmatter}


\section{Introduction}
IntraVascular UltraSound (IVUS) is a real-time tomographic imaging technology that visualizes the arterial wall by inserting an ultrasound transducer with a catheter into the artery and pulling back the transducer at a constant speed~\cite{4472859}. 
IVUS imaging is key to studying the vascular wall for the diagnosis and assessment of cardiovascular disease~\cite{4472859}, \cite{Nissen2001IntravascularUN}. 
For example, IVUS imaging is crucial for quantitative analysis of the vessel walls and atherosclerotic plaque characteristics and is also necessary for generating 3D reconstructed models of arteries~\cite{Faraji_2018}.

Generally, arteries are characterized by two distinct borders. 
The \emph{lumen} corresponds to the lumen-wall interface, and the \emph{media-adventitia} or \emph{media} corresponds to the media-wall~\cite{S0895611106001340}.
Segmentation of the lumen and media in IVUS images is the first step toward evaluating a vessel's morphology and identifying possible atherosclerotic lesions. 
However, the presence of speckles and various artifacts in the IVUS images, e.g., the existence of plaques, guide-wire artifacts, stents, shadow artifacts, bifurcations, and side vessels, increases the difficulty of lumen and media segmentation~\cite{Garcìa_Gogas_PMC3078312}.

Several automated and semi-automated approaches have been proposed to either jointly or individually segment lumen and media.
A review of earlier automatic segmentation techniques of IVUS images before the advent of neural networks can be found in~\cite{6159086}.
In \cite{9952588}, a Support Vector Machine (SVM) classification model is trained for lumen and media segmentation using the Gray-level co-occurrence matrix and Gabor features extracted from polar space-transformed images.
Lumen, media, and surrounding tissues are automatically detected using SVMs in \cite{LOVERCIO2019113}, where a Random Forest detects different morphological structures.
In~\cite{PMID_28624754}, the K-means algorithm with subtractive clustering is implemented for lumen segmentation, while in~\cite{Naceur_CEIT_2013}, a combined method for lumen segmentation is proposed.

Deep Neural Networks (DNNs) have been increasingly used for automatic segmentation of lumen and media boundaries of IVUS images~\cite{s13239_2023_arora}.
A multi-domain regularized deep learning segmentation method that leverages transfer learning from cross-domains, is proposed in~\cite{chen2016iterative}.
The work in \cite{BLANCO2022102262} relies on a DNN to deliver a preliminary segmentation, followed by a Gaussian process regressor to construct the final lumen and media contours. 
In \cite{9593822}, a coarse-to-fine strategy to facilitate image segmentation of lumen and media boundaries in IVUS images is proposed using the BASNet architecture~\cite{qin2021boundaryaware}.
An architecture with two sparse auto-encoders and one softmax classifier was proposed in~\cite{Zhang_CMIG_2017}, where the input images are first filtered through a mean filter and then the region of interest is extracted by connecting the brightest pixels in every radial direction. 

More recent techniques are based on the U-Net~\cite{ronneberger2015unet} architecture or its variants.
The extraction of lumen and media borders in IVUS images using a multi-scale feature-aggregation technique applying U-Net has been presented in \cite{9175970}.
In \cite{doi:10.1177/01617346221114137}, the segmentation of lumen and media is done by a feature pyramid network that was added to the U-Net++ model~\cite{zhou2018unet}, to enable the utilization of feature maps at different scales.
An architecture consisting of a VGG16-based encoder and a U-Net decoder is proposed in~\cite{simonyan2015deep} and it was shown to outperform its predecessors~\cite{balakrishna2018automatic}.
IVUS-Net~\cite{yang2018ivusnet} is another U-Net-like architecture for lumen and media segmentation giving state-of-the-art results. 
Both the IVUS-NET and the U-Net have an autoencoder based architecture.
The main difference between IVUS-Net and the basic U-Net is that in addition to the main branch, the IVUS-Net also has a “refining branch” for each encoder and decoder block that helps extract features at different scales.
Another segmentation method, SUMNET~\cite{8759210}, is inspired by U-Net and SegNet~\cite{badrinarayanan2016segnet}.
Compared to SegNet, SUMNET transfers not only the pooling indices but also the entire feature map in the encoder to the corresponding decoders. 
Finally, Kim \emph{et al.}~\cite{10.1007/978-3-030-01364-6_18} propose a multi-scale fully convolutional neural network for segmenting lumen and media from IVUS images. 
The main difference from U-Net is that this work has multi-scale inputs and outputs. 

The above DNN-based methods treat IVUS segmentation as a pixel segmentation problem and they classify each pixel into lumen, media, or background. 
As a result, they suffer from two important shortcomings.
Firstly, during training they optimize pixel-wise loss functions (with the most popular choice being cross-entropy).
On the other hand, for testing and performance evaluation they use the Jaccard measure (JM)\footnote{JM is the ratio of the area of the intersection between two closed curves, divided by the area of their union.} which is a more appropriate metric of the similarity between two image segments and is frequently used in medical image segmentation~\cite{9116807}.
However, it is well known that the loss function during training should be the same as the loss function during testing.
Secondly, the identified lumen and media regions often have very irregular shapes.
However, the shape of the lumen and media regions of the blood vessel resembles a conic section \cite{Faraji_2018},~\cite{PMID_24063959}. 
For that reason, it has been proposed to add a post-processing step that fits ellipses to the predicted pixel segmentation output~\cite{Faraji_2018}, \cite{NOBLE2020100042}, \cite{7897217}.
This extra step significantly improves the segmentation results but is computationally intensive and increases the latency in processing. 
Motivated by this, in our previous work~\cite{9231871}, we combined a DNN and an ellipse-fitting post-processing step into a single CNN-type architecture that directly predicted the parameters of an ellipse that best estimates the segmentation of Lumen and Media of IVUS images. 
Approximating the lumen and media boundaries by ellipses, either directly or via post-processing, does improve performance over IVUS segmentation methods that impose no constraints on the boundary shape. 
Nevertheless, we have observed that an elliptical shape constraint is still too simplistic and fails to capture the natural shape of the lumen and media boundaries.

In this paper, we extend our work in~\cite{9231871} by relaxing the shape constraint from an ellipse to a closed curve, specifically, a closed polygonal chain.
The closed polygonal chain is more flexible than an ellipse as it has more degrees of freedom and can model irregular shapes well.
The parameters of this closed polygonal chain are predicted with a subband-based CNN that has interesting and favorable properties that have been well-explored in~\cite{8804202} and~\cite{sinha2023structurally}.
We consider two loss functions when we train the CNN.
The first is a novel loss function that uses JM to quantify the similarities between the predicted lumen and media segments and the corresponding ground-truth image segments. 
We derive closed form expression for this loss function which we use to back-propagate the error derivatives and update the CNN weights. 
Work on employing JM as a loss function is seen in~\cite{wang2023jaccard}. They present a theoretical analysis of the properties of the JM and introduce a term called the Jaccard Metric Loss (JML) to compute the loss.
As for the second loss function, we use the Mean Squared Error (MSE) between the vertices of the predicted and the ground truth polygonal chains.  
The performance of the proposed method using either of the two loss functions is experimentally investigated.
We observe that the proposed JM Loss function mostly outperforms the MSE loss function for both JM and HD measures.

The paper is organized as follows. In Section \ref{sec:proposed_method}, we describe our proposed method and discuss its properties. Section~\ref{sec:Experimental setup and results} describes the experimental setup and results. Finally, we conclude this paper in Section \ref{sec:Conclusion}.

\begin{figure*}[t!]
    \centering
    \includegraphics[trim=0.9cm 0 0 0, scale=1.07]{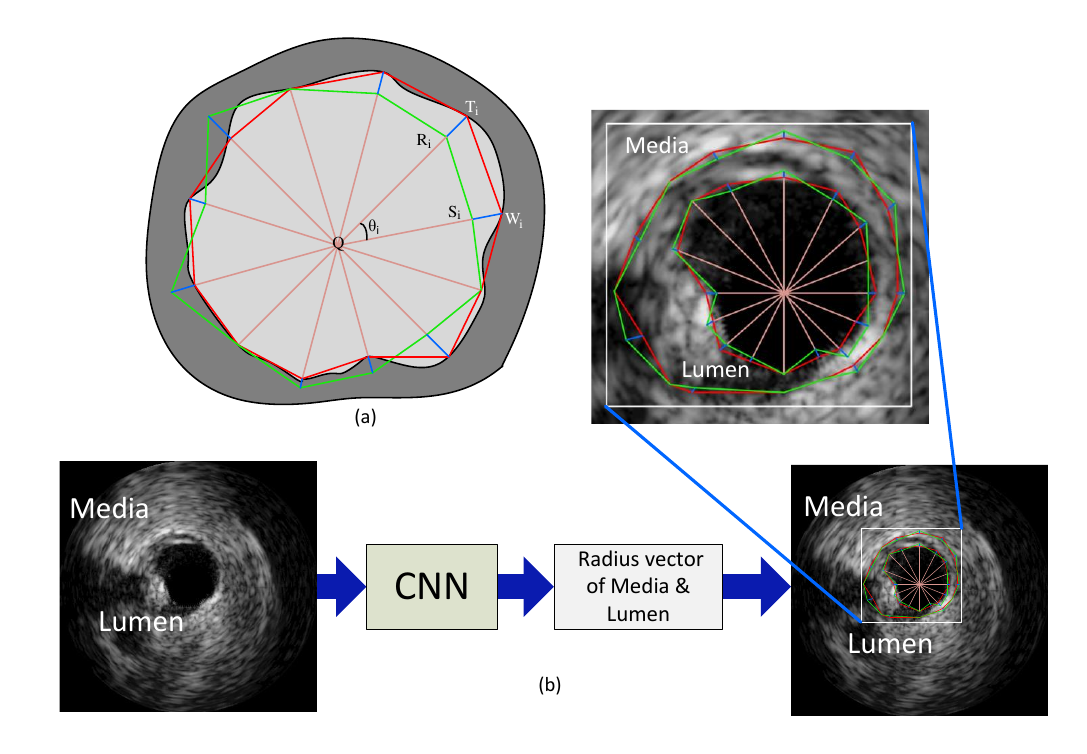}
    \caption{
    Figure (a) shows an example of the predicted closed curve (green) vs. ground truth (red). The difference in length between the predicted vertices and the ground truth points at every angle is marked in blue. 
    Figure (b) shows the system-level block diagram of the proposed method. The ground truths are represented by the red polygonal closed contour for both lumen and media, while the predicted curves are represented by the green polygonal closed contours for both lumen and media. 
    The input IVUS image is passed through a CNN that outputs two predicted vectors one for lumen and the other for media. Each vector is defined by 16 elements, which is also the $N_v$ of the model.
    The figure has been enlarged for better viewing.
    Both figures (a) and (b) are used for illustration purposes only.
    }
    \label{system_level_diagram}
\end{figure*}

\section{Proposed Method}
\label{sec:proposed_method}
Our method automatically segments the lumen and media boundaries by predicting the $N_v$ vertices of two closed polygonal chains, one for each of the two boundaries.
The vertices are described by their polar coordinates on the image plane originating at the center of the image.
Here we assume that the center of the image is close to the center of the lumen and media boundary contours.
Such an assumption is valid since the IVUS images are obtained using a catheter inserted through the middle of an artery or vein.
As we illustrate in Figure~\ref{system_level_diagram}(a), the vertices are placed at fixed angles obtained by dividing the entire 360\degree~angular space into equally spaced angles.
We obtain the ground truth polygonal curves for the lumen and media simlilarly, by sampling the boundaries at the predetermined fixed angles, and linearly joining the obtained points.
Hereafter, we refer to these points as `ground truth' points.
In Figure~\ref{system_level_diagram}(a), the red contour represents the ground truth, while the green contour represents the predicted contour; the blue line at each predicted point represents the delta between the predicted and the ground truth points.
We note that the process of sampling the ground truth boundaries at fixed angles to obtain the vertices of the ground-truth polygonal chains, introduces certain errors.
The analysis of this sampling error is beyond the scope of this work, and it can be reduced by increasing $N_v$.

Using a CNN, we predict the location of the vertices of the media and lumen closed contours. 
Figure~\ref{system_level_diagram}(b) shows the system-level block diagram of the proposed method. It shows the input image, a CNN predicting the vertices of both lumen and media, and an output image formed by joining the consecutive vertices of the two closed contours.
By fixing its angle, the location of a vertex is fully determined by its radius.
Therefore, the total number of components of the predicted output vector equals $2\times N_v$.
This approach lessens the burden on the CNN compared to the CNN predicting both the angle and radius.
Similarly, if the vertices were to be predicted in the Cartesian coordinate system, then each of the predicted vertices would consist of a pair of points for the x-and-y axis, and the CNN would have to predict both these values.

\begin{figure}[] 
    \centering
    \includegraphics[scale=0.65]{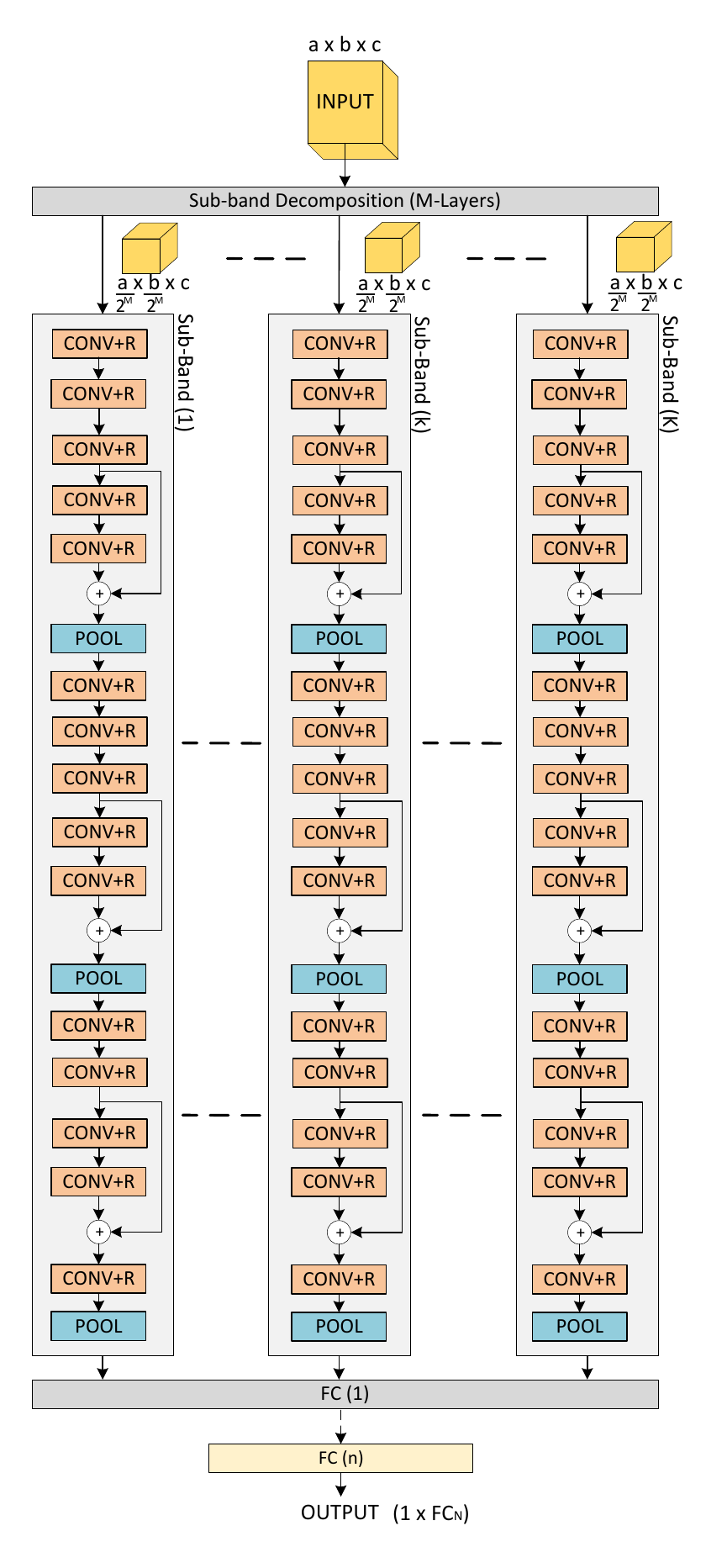}
    \caption{Structurally regularized  CNN architecture  via  Adaptive Subband Decomposition (ASD)~\cite{sinha2023structurally}, parametrized by input dimensions (a$\times$b$\times$c), number of subbands $K$, number of convolutional layers per subband $I$, number of FC layers $N$, and a pair of five output classes each representing an ellipse, one for lumen and the other for media. Since the input is a black and white image, the input channel contain the same values for red, green and blue colors channels. 
    }
    \label{CNN_arch}
\end{figure}
We use the Multi-Channel Subband Regularized CNN (MSR-CNN) Architecture of \cite{sinha2023structurally} shown in Figure~\ref{CNN_arch}. MSR-CNN has an Adaptive Subband Decomposition (ASD) frontend which decomposes the input image into subbands that are learnt from the dataset and then processes each of the subbands by an independent CNN that is significantly light weight. The outputs of these CNNs are combined by a fully connected layer providing the final output.

In our proposed method, the MSR-CNN directly predicts the lumen and media polygonal chains without requiring an encoder-decoder pair as in U-Net-based architectures.~\cite{ronneberger2015unet}.
The decoder in our architecture is reduced to simply drawing the predicted closed contours of the lumen and media segments by joining the consecutive predicted vertices. 
As such, our architecture drastically reduces computation cost as it bypasses the decoder and/or costly post-processing needed in the case of an auto-encoder-based solution that labels each pixel to lumen, media, or background~\cite{7950713}.

In addition,  as indicated in~\cite{sinha2023structurally}, the MSR-CNN architecture has certain properties that aid our design.
First, it provides robustness to input noise and weight quantization noise because the quantization error and noise of an individual subband are contained within that subband and do not affect the entire signal band. 
Second, the CNN structure provides regularization which is hugely needed for a small dataset like that of the IVUS dataset. 
Finally, the subband decomposition structure further reduces the overall computational cost. 

We consider two loss functions when we train the CNN: a novel loss function based on JM, and MSE.
The JM~\cite{PMID_24012215}, $JM(A,B)$, between two closed curves $A$ and $B$ is given by the ratio of the overlapping area of the two curves over the area of their union:
\begin{flalign}
JM(A,B) = \frac{|A \cap B|}{|A \cup B|}
\label{jaccard_eq_1}
\end{flalign}
The proposed loss function is given by
\begin{align}
\epsilon &= \epsilon_{\text{lumen}} + \epsilon_{\text{media}}\\
& = \sum_{i=1}^{N_v}\epsilon^{\text{lumen}}_i + \sum_{i=1}^{N_v}\epsilon^{\text{media}}_i\\
& = \sum_{i=1}^{N_v}(1-JM^{\text{lumen}}_i) + \sum_{i=1}^{N_v}(1-JM^{\text{media}}_i)
\end{align}
In the above, $JM^{\text{lumen}}_i$ ($JM^{\text{media}}_i$) is the JM between the $i$th predicted and ground truth lumen (media) angular segments.
Referring to Fig.~\label{system_level_diagram}(a), $JM_i$ is the JM between the triangles $Q-R_i-S_i$ and $Q-T_i-W_i$, where we dropped the superscript ``lumen''/``media'' to simplify the presentation.
The assumption here is that when we maximize the sum of JMs for each angular segment, this results in maximizing the overall JM for the entire lumen and media segments. 
\begin{figure}[t]
    \includegraphics[trim=0.9cm 0 0 0, scale=1.08]{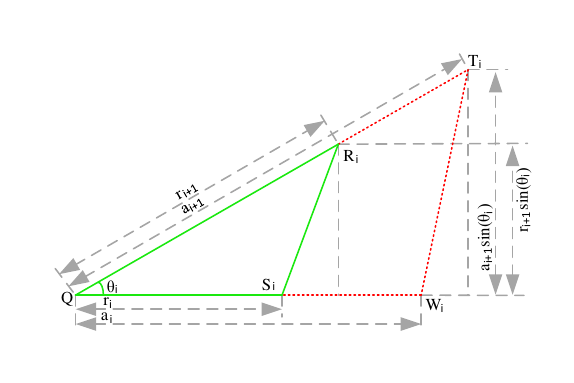}
    \caption{
    The Figure shows the slice Q-W2-T2 from Figure~\ref{system_level_diagram}(a). 
    The estimated triangle is indicated in green, while the ground truth is in red. 
    A section of the red triangle is overlapping with the green triangle, hence the red triangle is not fully visible. 
    This is only an example of many possible ways of overlap between the red and the green triangles.
    For a detailed description, please see Appendix-I. 
    We compute the JM between the red and the green triangle. 
    }
    \label{approx_exact_img}
\end{figure}


The exact form for $JM_i$ depends on the relative position of the points $R_i$, $S_i$, $T_i$ and $W_i$.
Overall, when we compute $JM_i$ there are six cases that we need to consider.
In \ref{sec:Appendix-I} we show the derivations for all six cases.
Here, for illustration purposes, we will discuss one such case shown in Figure~\ref{approx_exact_img}, where the green triangle indicates the predicted area, while the red triangle corresponds to the ground truth. 
The angle $\theta_{i}$ indicates the angle between the two arms of the triangle. 
The predicted length $r_{i}$, the ground truth distance $a_{i}$, and $a_{i+1}$ are also marked in the figure. 
Using trigonometric properties, we can calculate the lengths shown in the figure, using which we can see that the JM between the angular segments $Q-R_i-S_i$ and $Q-T_i-W_i$ is given by
\begin{flalign}
JM_i &= \frac{r_{i}r_{i+1}}{a_{i}a_{i+1}-r_{i}r_{i+1}}
\end{flalign} 
and the error is given by
\begin{flalign}
\epsilon_i &= 1 - JM_i = \frac{-a_{i}a_{i+1}}{r_{i}r_{i+1}-a_{i}a_{i+1}}
\end{flalign} 

The detailed computation of the JM 
and the back-propagation error for all possible overlapping triangular segments  are included in \ref{sec:Appendix-I}.

The second loss function we consider is the traditional MSE loss function commonly used in regression. 
In our case, it is the total error between the predicted and ground truth radii over all vertices for both lumen and media boundaries.
It is given by
\begin{align}
MSE &= \sum_{i=1}^{N_v}(r^{\text{lumen}}_i-a^{\text{lumen}}_i)^2 +
\sum_{i=1}^{N_v}(r^{\text{media}}_i-a^{\text{media}}_i)^2
\end{align}

\begin{figure*}[t]
    \centering
    \includegraphics[trim=6.5cm 0 0 0, scale=0.45]{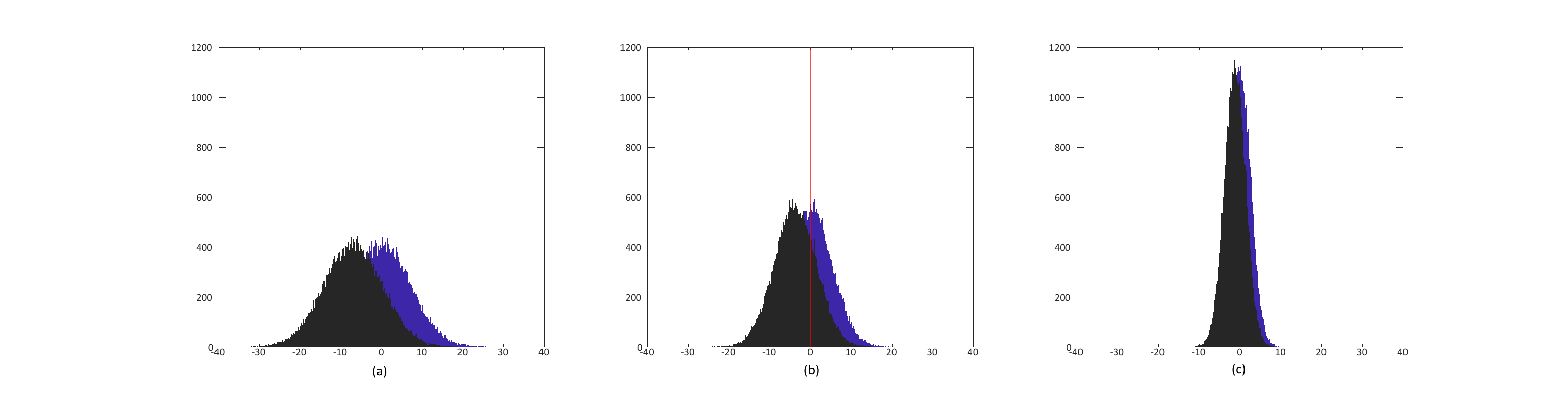}
    \caption{Histogram of Lumen and Media segmentation inference error between predicted points and ground truth computed in terms of number of pixels and their frequency of occurrence. (a) The blue histogram is with JM Loss and 16 $N_v$, while the black histogram is computed for the MSE loss with 16 $N_v$.
    (b) The blue histogram is with JM Loss and 32 $N_v$, while the black histogram is computed for MSE loss with 32 $N_v$.
    (c) The blue histogram is with JM Loss and 64 $N_v$, while the black histogram is computed for MSE loss with 64 $N_v$.
    We can observe that the histogram computed with the MSE method has a nonzero mean for each of the three model configurations, while the histogram computed with JM loss, almost has a zero mean. Further, all the histograms resemble Gaussian distribution for both JM and MSE loss functions. We have clubbed the error from both Lumen and Media predictions into the same bucket for error histogram analysis.
    }
    \label{result_hist}
\end{figure*}

\begin{figure*}[h!]
    \centering
    \includegraphics[trim=0.6cm 0 0 0, scale=0.87]{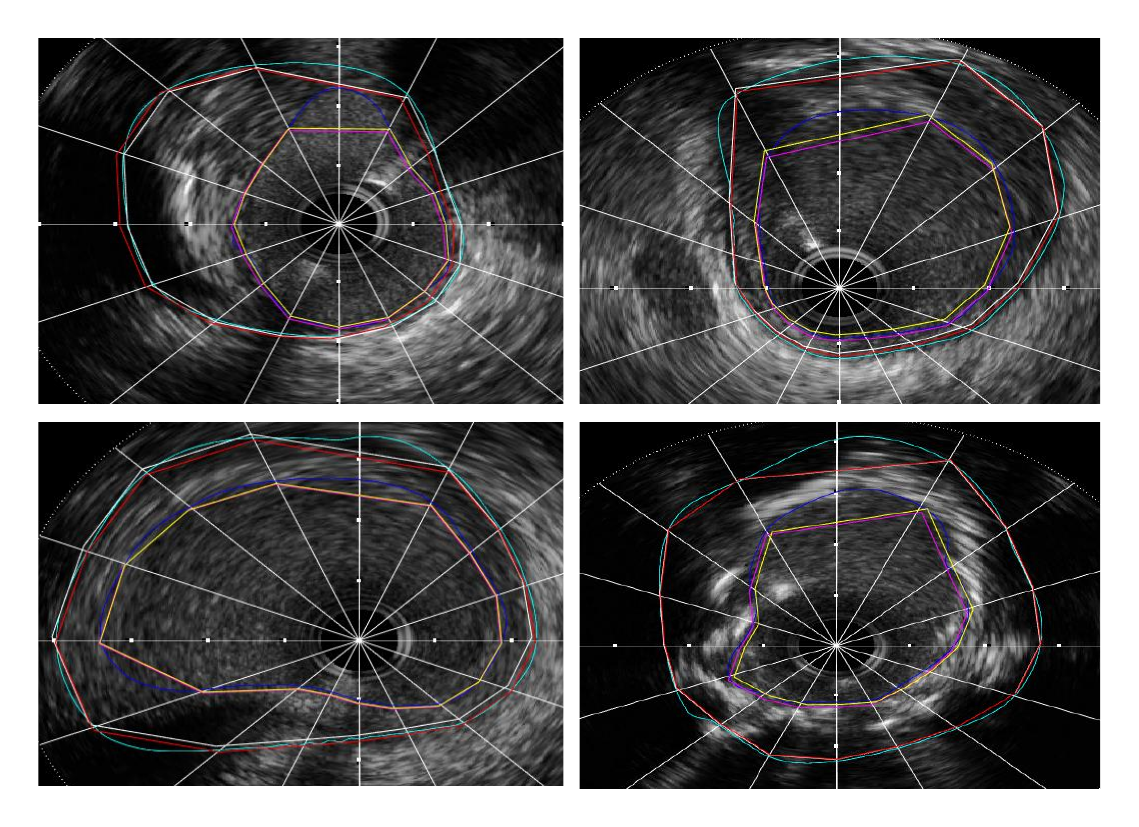}
    \caption{Lumen and media segmentation with 16 $N_v$. 
    Ground truth is marked by clinical experts indicating lumen and media are shown in blue and cyan, respectively.
    The predicted contours using the JM-Loss are indicated by magenta and white for Lumean and Media, respectively. 
    The predicted contour using the MSE-Loss is indicated by yellow and red for Lumean and Media, respectively.
    The four images have been chosen to have irregular contours to visually see the irregular contour prediction of Lumen and Media. Visually, we can see that the predicted contour closely resembles ground truth.}
    \label{result_images_1}
\end{figure*}

\begin{figure*}[h!]
    \centering
    \includegraphics[trim=0.6cm 0 0 0, scale=0.95]{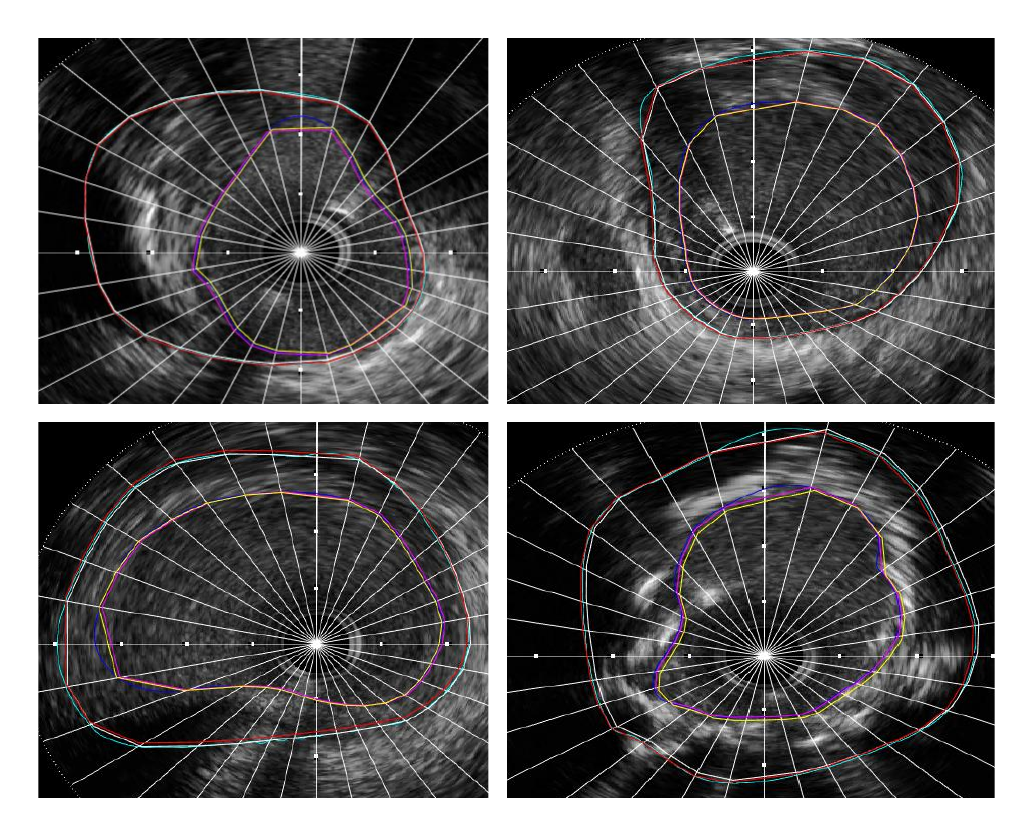}
    \caption{Lumen and media segmentation with 32 $N_v$. 
    Ground truth is marked by clinical experts indicating lumen and media are shown in blue and cyan, respectively.
    The predicted contours using the JM-Loss are indicated by magenta and white for Lumean and Media, respectively. 
    The predicted contour using the MSE-Loss is indicated by yellow and red for Lumean and Media, respectively.
    The four images have been chosen to have irregular contours to visually see the irregular contour prediction of Lumen and Media. Visually, we can see that the predicted contour closely resembles ground truth.}
    \label{result_images_2}
\end{figure*}

\section{Experimental Setup and Results}
\label{sec:Experimental setup and results}

\subsection{Methodology}
\subsubsection{Dataset}
We use the IVUS-Challenge-2011 dataset~\cite{PMID_24012215} from the University of Barcelona. 
Images in this dataset are IVUS gated frames using a full pullback at the end-diastolic cardiac phase from 10 patients.  
The labels in the IVUS dataset are provided as text files with x- and y-axis coordinate points marking the periphery of the lumen and media contours.
We scale the images using bi-linear interpolation to dimension $224\times 224\times 1$. 
The images of the IVUS dataset are greyscale, i.e., single channel.
We then replicate the single channel input into three channels to form the three input channels to the CNN, red, green, and blue, and thereafter represent the greyscale image in an RGB format.
It can be seen as a common practice to scale the images to resolution $224\times 224\times$ and using an RGB format replicating the single greyscale channel into three,~\cite{balakrishna2018automatic},~\cite{10135355},~\cite{8502126},~\cite{9231871}.
Since the data set contains only 109 images, we apply image augmentation to generate more data.
The data augmentation we implemented involves two steps. 
First, all images as well as their labels in the training set are: 
(1) flipped left to right; (2) flipped upside down; 
(3) flipped left to right then upside down; (4) rotated randomly from ${-}45~\text{deg}$ to $+45~\text{deg}$. Then, all generated data are corrupted by additive Gaussian noise with mean 0 and variance $0.2\times 255$.

\subsubsection{Proposed method}
We use the methodology described in~\cite{sinha2023structurally} to train the network.
At design time, the $N_v$ is determined, and the number of angles is fixed. 
Through our simulations, we found that an $N_v$ of 16, 32, and 64 provided competitive results in a reasonable amount of training time. 
For our study, we will study the models using various combinations of $N_v$ values, ASD decomposition architecture, and loss functions. Below is a list of the various model configurations chosen for comparison. 
\begin{enumerate}[label=(\roman*)]
\item The MSR-CNN architecture with ASD decomposition structure, $N_v$ set to 16, and JM loss function.
\item The MSR-CNN architecture with ASD decomposition structure, $N_v$ set to 16, and MSE loss function.
\item The MSR-CNN architecture with ASD decomposition structure, $N_v$ set to 32, and JM loss function.
\item The MSR-CNN architecture with ASD decomposition structure, $N_v$ set to 32, and MSE loss function. 
\item The MSR-CNN architecture with ASD decomposition structure, $N_v$ set to 64, and JM loss function.
\item The MSR-CNN architecture with ASD decomposition structure, $N_v$ set to 64, and MSE loss function.
\end{enumerate}

\subsubsection{Benchmarks}
We compare our models against seven benchmarks:
\begin{enumerate}[label=(\roman*)]
\item U-Net~\cite{ronneberger2015unet}.
\item IVUS-Net~\cite{yang2018ivusnet}. 
\item Inception+U-Net which is an auto-encoder network with Inception-v1~\cite{szegedy2014going} as the encoder and the decoder section from U-Net, followed by post-processing as described in~\cite{Faraji_2018}. 
\item Residual-Inception+U-Net which is similar to (iii) with the exception of adding residual connections to the Inception network.
\item GCN+U-Net which is similar to (iii), but it uses the GCN architecture in~\cite{8804202} for the auto-encoder.
\item IVUS-U-Net++~\cite{doi:10.1177/01617346221114137} which is a feature pyramid network added to the U-Net model, to enable the utilization of feature maps at different scales.
\item Subband CNN-wavelet~\cite{9231871} where a wavelet-based subband CNN is used to predict the ellipse parameters that best resemble the lumen and media curves.
\end{enumerate}



The U-Net-based architectures such as the IVUS-Net, and IVUS-U-Net++ or different derivatives of the U-Net architecture models produce a binary mask to classify the pixels as either lumen or media.
On the other hand,~\cite{9231871} uses a single model to generate both lumen and media boundary in a single run. 

\subsubsection{Training}

The computation platform used for evaluating the results is a Dell PowerEdge R740xd server with 256 GB DDR4 2666MT/S, dual Intel-Xeon Gold 5220 CPUs with 18 cores each, and an Nvidia GTX 1080Ti GPU with Nvidia CUDA and cuDNN libraries.
The Adam optimizer~\cite{kingma2017adam} is used with the parameters $\mu$, $\nu$ and $\epsilon$,  set to $0.9$, $0.999$ and ${10}^{-8}$, respectively, as suggested in~\cite{kingma2017adam}. 
A random 90 - 10 splitting strategy is used for training and validation, meaning that 90\% of the data before augmentation is randomly picked to train the models, and the remaining 10\% of the data is used for validation during the training process.
For each benchmark architecture, the models are trained to predict both lumen and media segmentation.
The proposed models are trained using the stochastic gradient descent over 10,000 epochs. We use a batch size of 32, batch normalized, randomly picked images per mini-batch, the momentum of 0.9, and weight decay of 0.0005~\cite{8804202}.
We update the weights as shown in~\cite{NIPS2012_4824}.
We initialize the learning rate to 0.01 and all biases to 1.
Initial weights are drawn from a Gaussian distribution with zero mean and a standard deviation of 0.01.

\begin{table*}[t]
\caption{ Comparison of JM and HD of the lumen and media segmentation.}
\centering
\label{jm_hd_dist}
\scalebox{0.8}
{
\begin{tabular}{|l|ll|ll|}
\hline
\multirow{2}{*}{}                                                                                                            & \multicolumn{2}{c|}{JM}                              & \multicolumn{2}{c|}{HD}                                \\ \cline{2-5} 
                                                                                                                             & \multicolumn{1}{l|}{Lumen}          & Media          & \multicolumn{1}{l|}{Lumen}           & Media           \\ \hline
U-Net~\cite{ronneberger2015unet}                                                                       & \multicolumn{1}{l|}{0.8642}         & 0.8355         & \multicolumn{1}{l|}{0.218}           & 0.3595          \\ \hline
IVUS-Net~\cite{yang2018ivusnet}                                                                        & \multicolumn{1}{l|}{0.8778}         & 0.8746         & \multicolumn{1}{l|}{0.21}            & 0.422           \\ \hline
\begin{tabular}[c]{@{}l@{}}Inception + U-Net + Post Processing~\cite{szegedy2014going}\end{tabular} & \multicolumn{1}{l|}{0.8364}         & 0.8344         & \multicolumn{1}{l|}{0.2848}          & 0.4638          \\ \hline
\begin{tabular}[c]{@{}l@{}}Res. Inception + U-Net + Post Processing~\cite{Faraji_2018}\end{tabular}     & \multicolumn{1}{l|}{0.8228}         & 0.832          & \multicolumn{1}{l|}{0.2968}          & 0.4759          \\ \hline
\begin{tabular}[c]{@{}l@{}}GCN + U-Net + Post Processing~\cite{8804202}\end{tabular}                 & \multicolumn{1}{l|}{0.854}          & 0.8576         & \multicolumn{1}{l|}{0.2389}          & 0.446           \\ \hline
IVUS-U-Net++~\cite{doi:10.1177/01617346221114137}                                                      & \multicolumn{1}{l|}{0.9412}         & 0.9509         & \multicolumn{1}{l|}{0.0639}          & 0.0867          \\ \hline
Subband CNN-Wavelet~\cite{9231871}                                                                     & \multicolumn{1}{l|}{0.895}          & 0.881          & \multicolumn{1}{l|}{0.217}           & 0.358           \\ \hline
\begin{tabular}[c]{@{}l@{}}MSR-CNN (16 $N_v$, JM-Loss)\end{tabular}                                                           & \multicolumn{1}{l|}{0.930}          & 0.915          & \multicolumn{1}{l|}{0.092}           & 0.098           \\ \hline
\begin{tabular}[c]{@{}l@{}}MSR-CNN (16 $N_v$, MSE-Loss)\end{tabular}                                                          & \multicolumn{1}{l|}{0.949}          & 0.938          & \multicolumn{1}{l|}{0.097}           & 0.099           \\ \hline
\begin{tabular}[c]{@{}l@{}}MSR-CNN (32 $N_v$, JM-Loss)\end{tabular}                                                           & \multicolumn{1}{l|}{0.961}          & 0.967          & \multicolumn{1}{l|}{0.0630}          & 0.0869          \\ \hline
\begin{tabular}[c]{@{}l@{}}MSR-CNN (32 $N_v$, MSE-Loss)\end{tabular}                                                          & \multicolumn{1}{l|}{0.958}          & 0.960          & \multicolumn{1}{l|}{0.0689}          & 0.0893          \\ \hline
\begin{tabular}[c]{@{}l@{}}MSR-CNN (64 $N_v$, JM-Loss)\end{tabular}                                                           & \multicolumn{1}{l|}{\textbf{0.969}} & \textbf{0.973} & \multicolumn{1}{l|}{\textbf{0.0601}} & \textbf{0.0843} \\ \hline
\begin{tabular}[c]{@{}l@{}}MSR-CNN (64 $N_v$, MSE-Loss)\end{tabular}                                                          & \multicolumn{1}{l|}{0.960}          & 0.965          & \multicolumn{1}{l|}{0.0657}          & 0.0861          \\ \hline
\end{tabular}
}
\end{table*}

\subsubsection{Performance metrics}
We quantify and compare the segmentation results using JM and HD~\cite{PMID_24012215}. 
HD is defined as the maximum distance between any pixel in the predicted segment contour and any pixel in the manual contour.
Consider two unordered, nonempty, discretized and bounced set of points $X$ and $Y$ and a distance metric $d(x,y)$ between the two points $x \in X$ and $y \in Y$; in our case, $d(\cdot,\cdot)$ is the Euclidean distance.
The HD between $X$ and $Y$ is defined in~\cite{ribera2019locating} as
\begin{flalign}
HD(X,Y) = \max_{x\in X,y\in Y} d(x,y)
\label{HD_equ_2}
\end{flalign}
As shown in~\cite{Attouch1991}, the HD is a metric. 
Also to note that during calculating the JM and HD measures, we compare the predicted closed contour against the original ground-truth points and not the sampled vertices.

\subsection{Results}
\label{sec:Results}
Figure~\ref{result_hist} shows the error histogram in terms of the difference in number of pixels between the predicted and the ground truth, clubbed together for lumen and media contours for a specific configuration of the model.
The histograms for both the JM loss and the MSE loss functions are shown in different colors.
We have combined the error for both lumen and media into a single bucket for the purpose of the histogram.
We can observe that the MSE loss function results in the predicted vertices having a bias in the histogram, while the error histogram from the JM loss function does not have any bias associated with it.
We see that both loss functions have a much tighter distribution when $N_v$ is increased. 
With 64 $N_v$, we observe that for both the error functions, it is within $\pm 5$ pixel errors. 
This shows that both methods give us significantly close results to the ground truth.

Table~\ref{jm_hd_dist} shows the segmentation results of the proposed and benchmark architectures. 
The bold values represent the best result for each metric (JM and HD). 
The proposed segmentation method with 64 $N_v$ and JM loss function, outperforms the benchmarks in both JM and HD metrics.
Importantly, the proposed architecture drastically reduces the network size by moving away from an auto-encoder architecture style, to just the encoder, and yet estimates the contours of lumen and media of any irregular shape that closely resembles the ground reality. 
Hence, we obtain both improved performance and a drastic reduction in computational resources.

Figure~\ref{result_images_1} shows the lumen and media segmentation for 16 $N_v$. The ground truth is marked by clinical experts indicating lumen and media are shown in blue and cyan, respectively. 
The predicted contours using the JM loss function are indicated by magenta and white, respectively. 
The predicted contour with the MSE loss function is indicated by yellow and red for the lumen and media, respectively. 
The four chosen images are among the difficult ones in the IVUS dataset and visually we can see that the predicted contour closely resembles ground truth.
Similar images have been used in Figure~\ref{result_images_2} and Figure~\ref{result_images_1} for visual comparison of the results between 16 $N_v$ and 32 $N_v$.
Our results closely match the ground truth even for irregularly shaped lumen and media contours.

\section{Conclusion}
\label{sec:Conclusion}
In this paper, we propose an automatic segmentation method of IVUS images for lumen and media contours.
The proposed method closely resembles the ground reality for contours that are highly irregular in nature.
We predict the vertices of the lumen and media contours. 
The predicted vertices are joined together to form the estimated closed contour lumen and media segments. 
In predicting the distances of each of the lengths corresponding to an angle, we calculate the JM and minimize it in each training iteration of the CNN. 
Increasing the number of predicted distances increases the number of angles to be predicted in the polar coordinate system and thereby increases the resolution of the predicted closed contour, which comes at a higher computational cost.
We use two loss functions, the JM, and the MSE loss function, and compare their results. 
We observe that both methods perform well, however, the JM loss function outperforms the MSE loss for each of the models.
We observe that as $N_v$ increases the predicted contour will eventually approach the ground truth contour. 
Our proposed method is simple to implement compared to other CNN algorithms that are based on autoencoder structures.
The decoder module in our method gets simplified to a function that simply joins the predicted vertices to form the closed contours of lumen and media. 
In each iteration during training, our algorithm increases the JM and thereby reduces the HD. 
Importantly, our method outperforms previous techniques even though they use a post-processing stage of ellipse-fitting on the pixel-by-pixel prediction of the lumen, media, and background regions.
Any of the known CNN networks could potentially be used in the encoder section to predict the vertices, however, the adaptive subband decomposition CNN provides better regularization which is an important factor when dealing with datasets like IVUS that contain very few images.
Our architecture produces results that are very close to the gold standard and outperforms most of the well-known architectures in literature, at a fraction of the computation cost.

\appendix
\section{Exact Jaccard Measure Computation}
\label{sec:Appendix-I}

\begin{figure*}[t]
    \centering
    \includegraphics[scale=0.67]{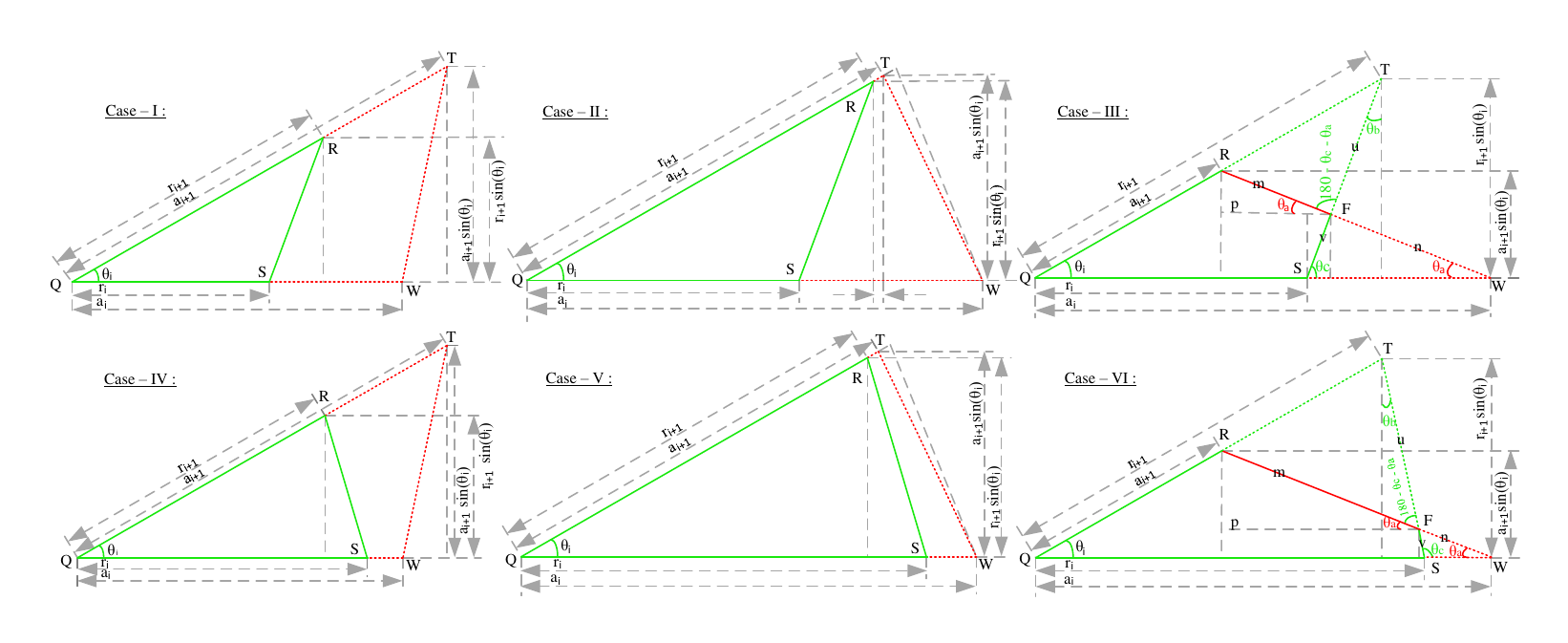}
    \caption{Six different scenarios for computing with the JM loss function. The green triangle indicates the predicted triangle, while the red triangle represents the ground truth. The length $r_{i}$ is the predicted length and the length $a_{i}$ is the ground truth. 
    The angle $/theta_{i}$ is the fixed angle chosen at design time. We divided the angle space into 16, 32, and 64 $T\__{p}$ for our models. 
    }
    \label{img_without_approx_i}
\end{figure*}

There are six cases to consider when computing with the JM loss function. These cases are shown in Figure~\ref{img_without_approx_i}.

Referring to Figure~\ref{img_without_approx_i} Case-I, $JM_{I}$ refers to the Jaccard Measure, $\epsilon_{I}$ refers to the error, $\frac{\partial \epsilon_{I}}{\partial r_{i}}$ is the partial derivative of the error with respect to the triangle arm $r_{i}$, $\frac{\partial \epsilon_{I}}{\partial r_{i+1}}$ is the partial derivative of the error with respect to the triangle arm $r_{i+1}$, for Case-I:
\begin{align}
JM_{I} &= \frac{\text{Num}}{\text{Den}} \\
\text{Num} &= \frac{1}{2}r_{1}^{2}\cos\theta_{i}\sin\theta_{i} - \frac{1}{2}(r_{i+1}\cos\theta_{i}-r_{i})r_{i+1}\sin\theta_{i} \\
\label{exact_case_2_JM_1_numerator}
\text{Den} &= \frac{1}{2}a_{i+1}^2\sin\theta_{i}\cos\theta_{i} - \frac{1}{2}(a_{i+1}\cos\theta_{i}-a_{i})a_{i+1}\sin\theta_{i} - \nonumber \\
&\frac{1}{2}r_{i+1}^2\sin\theta_{i}\cos\theta_{i} + \frac{1}{2}(r_{i+1}\cos\theta_{i}-r_{i})r_{i+1}\sin\theta_{i} \\
\label{exact_case_2_JM_1_denominator}
JM_{I} &= \frac{r_{i}r_{i+1}}{a_{i}a_{i+1}-r_{i}r_{i+1}}\\
\epsilon_{I} &= 1 - JM_{I} = \frac{-a_{i}a_{i+1}}{r_{i}r_{i+1}-a_{i}a_{i+1}}
\label{exact_case_1_JM_5}
\end{align} 
\\
Referring to Figure~\ref{img_without_approx_i} Case-II, $JM_{II}$ refers to the Jaccard Measure, $\epsilon_{II}$ refers to the error, $\frac{\partial \epsilon_{II}}{\partial r_{i}}$ is the partial derivative of the error with respect to the triangle arm $r_{i}$, $\frac{\partial \epsilon_{II}}{\partial r_{i+1}}$ is the partial derivative of the error with respect to the triangle arm $r_{i+1}$, for Case-II:
\begin{align}
JM_{II} &= \frac{\text{Num}}{\text{Den}} \\
\label{exact_case_2_JM_2_numerator}
\text{Num} &= {\frac{1}{2}r_{i}r_{i+1}\sin\theta_{i}} \\
\label{exact_case_2_JM_2_denominator}
\text{Den} &= \frac{1}{2}r_{i+1}\sin\theta_{i}(r_{i+1}\cos\theta_{i}-r_{i}) + \nonumber \\
&\frac{1}{2}a_{i+1}\sin\theta_{i}(a_{i}-a_{i+1}\cos\theta_{i}) + \nonumber \\ 
&\frac{1}{2}(a_{i+1}-r_{i+1})^2\sin\theta_{i}\cos\theta_{i} + \nonumber \\
&\frac{1}{2}r_{i+1}(a_{i+1}-r_{i+1})\cos\theta_{i}\sin\theta_{i}
\end{align} 

{Combining~\ref{exact_case_2_JM_2_numerator} and~\ref{exact_case_2_JM_2_denominator}} we get
\begin{align}
JM_{II} &= \frac{-r_{i}r_{i+1}}{r_{i}r_{i+1}-a_{i}a_{i+1}} \\
\epsilon_{II} &= 1 - JM_{II} = \frac{2r_{i}r_{i+1}-a_{i}a_{i+1}}{r_{i}r_{i+1}-a_{i}a_{i+1}}
\label{exact_case_2_JM_5}
\end{align} 
\\
Referring to Figure~\ref{img_without_approx_i} Case-III, $JM_{III}$ refers to the Jaccard Measure, $\epsilon_{III}$ refers to the error, $\frac{\partial \epsilon_{III}}{\partial r_{i}}$ is the partial derivative of the error with respect to the triangle arm $r_{i}$, $\frac{\partial \epsilon_{III}}{\partial r_{i+1}}$ is the partial derivative of the error with respect to the triangle arm $r_{i+1}$, for Case-III:
\begin{align}
a + b &= \frac{a_{i}-a_{i+1}\cos\theta_{i}}{\cos\theta_{ia}} = K_{1} \\
\label{exact_case_3_JM_1}
c + d &= \frac{r_{1}\sin\theta_{i}}{\cos\theta_{ib}} \\
\label{exact_case_3_JM_2}
P &= a\cos\theta_{ia}\\
\label{exact_case_3_JM_3}
d\sin\theta_{ic} &= b\sin\theta_{ia} \\
P &= d\cos\theta_{ic}+(r_{i} - a_{1}\cos\theta_{i0})
\label{exact_case_3_JM_5}
\end{align} 

Combining \ref{exact_case_3_JM_2} and \ref{exact_case_3_JM_5} we get
\begin{align}
\label{exact_case_3_JM_6} 
d &= \frac{a\cos\theta_{ia}+a_{i+1}\cos\theta_{i0}-r_{i}}{\cos\theta_{ic}}
\end{align} 

Combining~\ref{exact_case_3_JM_3}~and~\ref{exact_case_3_JM_6} we get
\begin{align}
b &= \frac{d\sin\theta_{ic}}{\sin\theta_{ia}}=\frac{(a\cos\theta_{ia}+a_{i+1}\cos\theta_{i0}-r_{i})}{\cos\theta_{ic}}\frac{\sin\theta_{ic}}{\sin\theta_{ia}}
\label{exact_case_3_JM_7}
\end{align} 

Combining~\ref{exact_case_3_JM_1}~and~\ref{exact_case_3_JM_7} we get
\begin{align}
a+b &= K_{1} = \frac{\text{Num}}{\text{Den}} \\
\text{Num} &= a\cos\theta_{ic}\sin\theta_{ic}+a\cos\theta_{ia}\sin\theta_{ic}+ \nonumber \\ &a_{i+1}\cos\theta_{i0}\sin\theta_{ic}-r_{i}\sin\theta_{ic} \\
\text{Den} &= \cos\theta_{ic}\sin\theta_{ia}
\end{align}
\begin{align}
a &= \frac{r_{i}\sin\theta_{ic}+(K_{1}\cos\theta_{ic}\sin\theta_{ia}-a_{i+1}\cos\theta_{i0}\sin\theta_{ic})}{\cos\theta_{ic}\sin\theta_{ia}+\cos\theta_{ia}\sin\theta_{ic}}
\label{exact_case_3_JM_8} 
\end{align} 
Therefore, we get
\begin{align}
b &= K_{1} - a \\
d &= \frac{a\cos\theta_{ia}}{\cos\theta_{ic}} + \frac{a_{i+1}\cos\theta_{i0}-r_{i}}{\cos\theta_{ic}} \\
d &= r_{i}K_{2} + K_{3}\\
c &= r_{i+1}K_{4} - r_{i}K_{2}-K_{3}\\
P &= a\cos\theta_{ia} \\
\label{exact_case_3_JM_9}
JM_{III} &= \frac{\text{Num}}{\text{Den}} \\
\text{Num} &= \frac{1}{2}a_{i+1}^2\cos\theta_{i0}\sin\theta_{i0} + \frac{1}{2}Pa\sin\theta_{ia} + \nonumber \\ &Pd\sin\theta_{id} - \frac{1}{2}d^2\cos\theta_{id}\sin\theta_{id}\\
\text{Den} &= \frac{1}{2}(a_{i} - r_{i})d\sin\theta_{ic} + \frac{1}{2}c.a\sin(\pi-(\theta_{ic} + \theta_{ia}))\\
\epsilon_{III} &= 1 - JM_{III} 
\end{align} 
\\
Referring to Figure~\ref{img_without_approx_i} Case-IV, $JM_{IV}$ refers to the Jaccard Measure, $\epsilon_{IV}$ refers to the error, $\frac{\partial \epsilon_{IV}}{\partial r_{i}}$ is the partial derivative of the error with respect to the triangle arm $r_{i}$, $\frac{\partial \epsilon_{IV}}{\partial r_{i+1}}$ is the partial derivative of the error with respect to the triangle arm $r_{i+1}$, for Case-IV:
\begin{align}
JM_{IV} &= \frac{\text{Num}}{\text{Den}} \\
\text{Num} &= r_{i}r_{i+1}\sin\theta\\
\text{Den} &= a_{i+1}^2\sin\theta\cos\theta - 
r_{i}r_{i+1}\sin\theta- \nonumber \\ 
&~~~~a_{i+1}\sin\theta(a_{i+1}\cos\theta-a_{i})\\
JM_{IV} &= \frac{-r_{i}r_{i+1}}{r_{i}r_{i+1}-a_{i}a_{i+1}} \\
\epsilon_{IV} &= 1 - JM_{IV} = \frac{2r_{i}r_{i+1}-a_{i}a_{i+1}}{r_{i}r_{i+1}-a_{i}a_{i+1}}
\label{exact_case_4_JM_3}
\end{align} 
\\


Referring to Figure~\ref{img_without_approx_i} Case-V, $JM_{V}$ refers to the Jaccard Measure, $\epsilon_{V}$ refers to the error, $\frac{\partial \epsilon_{V}}{\partial r_{i}}$ is the partial derivative of the error with respect to the triangle arm $r_{i}$, $\frac{\partial \epsilon_{V}}{\partial r_{i+1}}$ is the partial derivative of the error with respect to the triangle arm $r_{i+1}$, for Case-V:
\begin{align}
JM_{V} &= \frac{\frac{1}{2}r_{i}r_{i+1}\sin\theta}{\frac{1}{2}a_{i}a_{i+1}\sin\theta - \frac{1}{2}r_{i}r_{i+11}\sin\theta} \nonumber \\
\label{exact_case_5_JM_1}
&= \frac{-r_{i}r_{i+1}}{r_{i}r_{i+11}-a_{i}a_{i+1}} \\
\epsilon_{V} &= 1 - JM_{V} = \frac{2r_{i}r_{i+1}-a_{i}a_{i+1}}{r_{i}r_{i+1}-a_{i}a_{i+1}}
\label{exact_case_5_JM_3}
\end{align} 
\\

Referring to Figure~\ref{img_without_approx_i} Case-VI, $JM_{VI}$ refers to the Jaccard Measure, $\epsilon_{VI}$ refers to the error, $\frac{\partial \epsilon_{VI}}{\partial r_{i}}$ is the partial derivative of the error with respect to the triangle arm $r_{i}$, $\frac{\partial \epsilon_{VI}}{\partial r_{i+1}}$ is the partial derivative of the error with respect to the triangle arm $r_{i+1}$, for Case-VI. 
Equations (\ref{exact_case_3_JM_1}) to (\ref{exact_case_3_JM_9}) map exactly the same.
\begin{align}
JM_{VI} &= \frac{\text{Num}}{\text{Den}} \\
\text{Num} &= \frac{1}{2}a_{i+1}^2\cos\theta_{i0}\sin\theta_{i0}+\frac{1}{2}Pa\sin\theta_{ia}+\nonumber \\
&Pd\sin\theta_{id}-\frac{1}{2}d^2\cos\theta_{ic}\sin\theta_{ic} \\
\text{Den} &= \frac{1}{2}(a_{i}-r_{i}-d\cos\theta_{ic})d\sin\theta_{ic}+ \nonumber \\
&\frac{1}{2}ca\sin(\pi-(\theta_{ic}+\theta_{ia})) \\
\epsilon_{VI} &= 1 - JM_{VI} 
\label{exact_case_6_JM_2}
\end{align} 
\\

\bibliographystyle{model2-names.bst}\biboptions{authoryear}
\bibliography{main}

%

\end{document}